\documentstyle[12pt,fleqn,cite]{article}
\newcommand{\sect}[1]{\setcounter{equation}{0}\section{#1}}
\newcommand{\subsect}[1]{\subsection{#1}}
\newcommand{\subsubsect}[1]{\subsubsection{#1}}

\def\sxn#1{\sect{#1}}
\def\subsxn#1{\subsect{#1}}
\def\subsubsxn#1{\subsubsect{#1}}

\def\3mat#1#2#3#4#5#6#7#8#9 {\left(
\begin{array}{ccc}#1 & #2 & #3 \\ 
#4 & #5 & #6 \\ #7 & #8 & #9 \end{array}\right)}
\def\2mat#1#2#3#4{\left
(\begin{array}{cc}#1 & #2 \\ 
#3 & #4 \end{array}\right)}

\def\be{\begin{equation}}
\def\ee{\end{equation}}
\def\bea{\begin{eqnarray}}
\def\eea{\end{eqnarray}}

\begin{document}
\begin{titlepage}
\begin{center}
\vskip .2in

{\Large \bf U-duality and Network Configurations of Branes}
\vskip .5in

{\bf Alok Kumar and Subir Mukhopadhyay\\
\vskip .1in
{\em Institute of Physics,\\
Bhubaneswar 751 005, INDIA}}
\end{center}

\begin{center} {\bf ABSTRACT}
\end{center}
\begin{quotation}\noindent
\baselineskip 10pt
We explicitly write down the invariant supersymmetry conditions for 
branes with generic values of moduli and U-duality 
charges in various space-time dimensions $D \leq 10$. 
We then use these results to obtain new BPS states, 
corresponding to network type structure of such branes. 
\end{quotation}
\end{titlepage}
\vfill
\eject


\sxn{Introduction}

String\cite{sch,mukhi,sen,zwib,rey,lee,mats,berg,
bhatt,cal,rest,bk,rovel,smo2,smolin} 
and Brane\cite{wit,aha,leung} 
Networks have recently attracted a great deal of 
attention in various contexts. In the realm of string
theory, the fact that they come out as stable 
non-perturbative BPS states implies that a complete
knowledge of these states and their dynamics will 
make our understanding of non-perturbative regime
more transparent. More optimistically, it has been 
suggested, that it may be possible to construct a 
non-perturbative formulation of string theory by
treating these configuration as fundamental objects.
In fact, a non-perturbative, background independent
formulation for the dynamics of these networks has already
been proposed, albeit in an utterly different 
context\cite{rovel,smo2,smolin}.
Among other aspects, these networks have also made their 
appearances in connection with certain non-trivial 
compactifications\cite{leung}.
More precisely, when M-theory is considered on a 
toric geometry, such configurations appear along the locus 
of the vanishing cycle. 
Moreover these networks have also been used 
for understanding the symmetry enhancements in F-Theory
from type IIB string theory point of view\cite{zwib}. 
In this context,
it has been pointed out that the states corresponding to 
string junctions and nets are responsible for 
the symmetry enhancement
to the exceptional groups in F-Theory. 
 
The networks are also intimately connected to gauge 
theories due to the fact that the branes, in the weak 
coupling limit, get decoupled from the bulk supergravity
theory and corresponding low energy effective dynamics is described
by SYM gauge theories. These connections predict
non-perturbative states in gauge theories.
In particular, a finite network with transverse branes at the
ends gives rise to non-perturbative states in the world-volume
gauge theories of the boundary branes. Conversely, the positions
of the branes at the boundary arise as moduli in the gauge theory of 
the network. By varying these and other moduli, including 
those for internal branes, one can study the 
non-perturbative aspects of the theory like the coulomb and 
the Higgs branch and phase transitions between them.
Similar kinds of networks have also been found to play crucial roles
in various other contexts 
like quantum gravity and branched polymer\cite{smo2}.

Most of the discussions of the networks, however, 
involve strings and branes in
ten dimensions. In particular, in type IIB 
string theory, there exists $SL(2, Z)$ 
S-duality group and various supersymmetric networks 
with S-duality charges have been constructed by making 
use of it. However, it is well
known that compactified theories in lower dimensions contain various 
branes which are not present in the ten dimensional theories.
In addition to that, in lower dimensions, the $U$-duality group is 
much bigger and therefore can accommodate various new supersymmetric
configurations. In view of this, and keeping in mind
the varied applications of 
networks, it is interesting and useful to consider networks 
in lower dimensions which
will provide deeper insights in the domain of non-perturbative
physics. 

In this paper, we give a detailed analysis of 
brane configurations with U-duality charges. More precisely,
we write down the U-duality invariant conditions on 
supersymmetry transformation paramaters for various  
branes with U-duality charges, referred as
U-branes, in type II theories in space-time dimensions 
$D\leq 10$. Although some of  these results are implicit 
in the supersymmetry properties of branes discussed in other 
contexts \cite{cvet,pope,malda}, the network construction 
requires an explicit expression for the supersymmetry 
charges or, alternatively, the corresponding
parameters. We write down these expressions for 
a number of examples. 
To argue that such conditions are indeed true, we point out that  
brane solutions, with generic U-duality charges, 
can be obtained from some 
specific configuration by U-duality transformations.  
The condition on the supersymmetry parameters
are then also generated by the same tranformation.  
The consistency of such conditions on supersymmetry 
parameters is also seen from the fact that by setting
all the moduli to zero, they reproduce the supersymmetry 
condition for an appropriate brane configuration in 
ten dimensional type IIA/IIB theory and eleven dimensional
M-theory. Moreover, our results, in several examples, 
are also consistent with a classification of U-branes in 
a different context\cite{malda}.  

We then use the above expressions of the preserved supersymmetry 
charges to construct several new non-perturbative BPS configurations. 
It is observed that the supersymmetry condition for 
a U-brane involves a map from the physical
space to the internal space of $U$-duality. This suggests that one 
can construct a supersymmetric configuration of 
arbitrary number of branes, corresponding to a network structure,
provided they are oriented in the physical space 
appropriately\cite{bhatt}.
These configurations preserve one-quarter or less supersymmetries.

The plan of the paper is as follows. 
We start by reviewing the supersymmetries of the
brane configurations in ten dimensions in the next section. 
In section-3 we obtain the explicit form of the 
supersymmetries of U-branes 
using the transformations of fields and supercharges. We will consider
all lower dimensions down to four and will restrict ourselves to the 
toroidal compactification of type IIA/B theories. In 
section-4 we will present various brane networks that preserve
a certain amount of supersymmetry. 
We will conclude with the discussions of the results and
comments about further work.
A sketch for the derivation of the supersymmetry conditions
is presented in an appendix.


\sxn{Branes in ten dimensions}



To start with, we review the situation in ten
dimensional type IIB theory and present the results
for the supersymmetry of various brane
configurations. As is well known, the massless sector
contains a complex scalar $\tau$, two 2-form fields 
$B^{(1)}$ and $B^{(2)}$ and a self-dual 4-form field $A$.
The $U$ or $S$-duality group 
is $SL(2)$ with the maximal compact
subgroup being $U(1)$. The antisymmetric tensor fields transform
under $SL(2)$, the supercharges transform under $U(1)$ and
the complex scalar $\tau$ which parametrizes the coset $SL(2)/U(1)$
transform under both.

In order to give the explicit transformation let us introduce
the $2\times 2$ upper triangular vielbein matrix
${1\over\tau_2}\2mat{\tau_1}{\tau_2}{0}{1} = V_i^a(\tau)$
where $i$ and $a$ are the $SL(2)$ and $SO(2)$ indices 
respectively. Under a generic $SL(2)$ element
$\Lambda=\2mat{a}{b}{c}{d}$ it transform as
\be
V(\tau) \longrightarrow V(\tau') = \Lambda V(\tau) {\cal O}^{-1},
\ee
where ${\cal O}$ is the corresponding $SO(2)$ element
with angle $\arg(c\tau+d)$.
One can then easily check that under this transformation, 
\begin{displaymath}
\tau \longrightarrow {a\tau + b \over c\tau + d }.
\end{displaymath}
The self-dual 4-form field is a singlet
while the 2-form fields transform as
\be
\left(\begin{array}{c} B^{(1)} \\ B^{(2)} \end{array} \right)
\longrightarrow 
\2mat{a}{-b}{c}{-d}
\left(\begin{array}{c} B^{(1)} \\ B^{(2)} \end{array} \right).
 \label{B10d} \ee 
Finally, the parameters of supersymmetry 
transform as
\be
(\epsilon_L - i\epsilon_R )\rightarrow
             \exp({i\over2} arg(c\tau + d ))
             (\epsilon_L - i\epsilon_R ).
 \label{epsa10d} \ee

Now we consider branes with generic $SL(2)$ charges.
Type IIB theory allows only odd $p$-branes and every
$p$-brane couples electrically (or magnetically) to $(p+1)$-
form (or $(7-p)$-form) field. So there are instantons (7-branes)
coupled to scalars, strings (5-branes) coupled to $B$
and dyonic 3-brane coupled to $A$. The condition on the 
supersymmetry parameter can now be written down from 
the symmetry and charges. 

\setcounter{subsection}{1}
\subsubsxn{Strings and five branes}

Let us start with the string. The supersymmetric configurations 
involving $(p, q)$ strings and 5-branes have already been studied 
in some detail in literature \cite{sen,aha}.
However we review them for completeness. For a $(1, 0)$ 
or a D-string, the supersymmetry parameters mentioned above 
satisfy the condition:
\be
        \epsilon_L = \Gamma_{1..8}\epsilon_R.
                                \label{dstring}
\ee 
This can be complexified into a form:
\be
        \epsilon_L + i \epsilon_R = 
                i \Gamma_{1..8} (\epsilon_L - i \epsilon_R).
                                \label{10d1}
\ee 
The $SL(2)$ invariant form of the above equation 
is given by\cite{sen}
\be
        \epsilon_L + i \epsilon_R = e^{i \alpha} \Gamma_{1..8} 
                        (\epsilon_L - i \epsilon_R),
                        \label{10d1pq}
\ee
where $\alpha$ is a phase defined in terms of the $SL(2)/U(1)$ 
moduli $\tau$ and the $SL(2)$ charges $(p, q)$ as:
$p + q \tau  = |p + q \tau| e^{i \alpha}$. 

This $SL(2)$ invariant relation can be obtained
by starting from the condition for $(1,0)$ string with trivial 
modulus $\tau=i$ and making an $SO(2)$ transformation ${1\over
\sqrt{p^2+q^2}}\2mat{p}{-q}{q}{p} $ which takes $(1, 0)$ to
$(p, q)$ but keeps $\tau$ invariant. 
Then to get the nontrivial moduli one
has to make another $SL(2)$ transformation by the element
${1\over \tau_2}\2mat{\tau_1}{\tau_2}{0}{1}$ 
which leads to the above condition \cite{sl2}.

A similar construction is also possible for 5-branes. 
One can start with the supersymmetric condition for the D-5 brane
along $(5\cdots 9)$ hyperplane with trivial moduli:
\be
        \epsilon_L + i \epsilon_R = 
        i \Gamma_{1..4} (\epsilon_L - i \epsilon_R),
\label{d510d}\ee
and then make the transformations to get the $(p, q)$
5-brane with general moduli.
An individual $(p, q)$ 5-brane preserves a supersymmetry which is
similar to the one in (\ref{10d1pq}) and has a form:

\be
        \epsilon_L + i \epsilon_R = e^{i \alpha} 
         \Gamma_{1..4} (\epsilon_L - i \epsilon_R).
 \label{10d5pq} \ee

These supersymmetry conditions have the interesting implication 
that one can construct networks
of strings or 5-branes which break one-quarter 
supersymmetry and thus lead 
to stable BPS states. As has been shown in 
literature, one can consider any
number of planar $(p,q)$ strings provided 
their orientations are parallel
to the orientations of the respective charge 
vectors in a charge plane. Similarly,
the web or net of $(p, q)$ 5-branes
can be constructed  by having four common 
directions for each of the 5-brane
and by aligning one of the edges of each of 
them in a plane, similar to 
the case of $(p, q)$ string.

\subsubsxn{Three Brane}

The D3-brane couples to the self-dual 4-form 
field which is invariant under the $S$-duality. 
The supersymmetry parameters satisfy the condition: 
\be     \epsilon_L + i \epsilon_R = -i \Gamma_{1..6}
                        (\epsilon_L + i \epsilon_R).
                                \label{10d3}
\ee
The supersymmetry properties of
configurations of 3-branes therefore depend only on
their orientations in space. Though the 3-branes 
by themselves can not be used to construct  
non-trivial junctions, they do play an important role
in the analysis of junctions and webs.

\subsubsxn{Seven Brane and Instanton}


Finally we consider the 7-brane which is magenetically
charged with respect to the scalar fields \cite{bbg}. The
supersymmetry condition for a D-7 brane along
$(3,4,\cdots 9)$ is given by $\epsilon_L = \Gamma_{12}
\epsilon_R$. By writing this relation as 
\be
        \epsilon_L + i \epsilon_R = 
        - i \Gamma_{1 2} (\epsilon_L + i \epsilon_R),
                                \label{10d7}
\ee
we observe that it is manifestly $S$-duality invariant and 
therefore represents the supersymmetry condition 
for a general $(p, q)$ 7-brane as well. It is noted that a 
generic $(p, q)$ brane preserves a supersymmetry which is
independent of both, charges and moduli. 

The results of the above paragraph shows that any number of
parallel 7-branes of arbitrary (p, q) charges can be
introduced without breaking further supersymmetry. 
This is consistent with the analysis of BPS states in
F-Theory using string junctions. It has 
been pointed out\cite{zwib} that the
exceptional groups observed in F-Theory are reproduced 
from the type IIB point of view
by inroducing $(p, q)$ branes for $p, q \neq 1, 0$, 
and by analysing the states
constructed by connecting these branes. We notice 
that all such branes are parallel to
each other, irrespective of the values of $p$ and $q$.  

The case of $(p, q)$ instanton \cite{bbg} in type IIB
theory is similar, as they have similar $SL(2)$ 
properties as that of a 7-brane. One has to take 
euclidean signature
and replace $\Gamma_{12}$ by unity.

\sxn{Supersymmetry conditions for $U$ branes in lower dimensions}


\subsxn{Nine dimension}


Now we consider the type II theory compactified on an $S^1$.
The massless spectrum contains 
3 scalar fields parametrizing the coset $GL(2,R)/SO(2)$ \cite{ss}.
The other antisymmetric fields are three 1-form, two 2-form and one 
3-form fields. So in this case we will have all the branes
starting from instanton to 6-branes.

The $U$-duality group is $SL(2, Z)\times Z_2$ which is the discrete
subgroup of the supergravity duality group $SL(2,R)\times O(1,1)$.
The $SL(2)$ is identical to the S-duality in ten dimensions.
Under $SL(2)$, two 1-form fields transform as a doublet
and one as a singlet. The 2-forms transform 
as a doublet and the 3-form is a singlet.
Two of the three scalars 
form the coset $SL(2)/SO(2)$. Their transformations, like
in ten dimensions, can
be written in terms of $SL(2)/SO(2)$ vielbein. 
However there is an extra scalar coming from the volume
of the $T^2$ 
which remains singlet under $SL(2)$. The supersymmetry
parameters $\epsilon_{\pm}$ transform as a spinor 
of $SO(2)$. In the following 
analysis we concentrate on the $SL(2)$ part of the duality 
symmetry only.

\subsubsxn{Strings and Four branes}

The strings and the 4-branes couple electrically and magnetically
respectively to the 2-form field. From the M-theory point of view 
they are precisely the M2-branes and M5-branes wrapped on 1-cycles
of $T^2$ and form a doublet. On the other hand, from type IIB side
the strings are the same $(p,q)$ strings of ten dimension while
4-branes are the $(p,q)$ 5-branes wrapped on $S^1$.

Since the 2-form fields transform as a doublet under $SL(2)$,
we will have a doublet of strings and 4-branes.
Following the discussion in ten dimensions, we can write the 
$SL(2)$ invariant
supersymmetric condition for a string along $X^8$ as:
\be
       \left(\begin{array}{c} \epsilon_+ \\
            \epsilon_- \end{array}\right)
        = i\2mat{0}{e^{i \alpha}}{e^{-i \alpha}}{0}
        \Gamma_{1..7} 
        \left(\begin{array}{c} \epsilon_+ \\
           \epsilon_- \end{array}\right),
                        \label{9d1pq}
\ee
where $\epsilon_+$ and $\epsilon_-$ are the 
spinors in nine dimensions and $\alpha = \arg(p+q\tau)$.
Similarly, for the 4-branes, one can write the 
supersymmetry conditon in the same manner. One has 
to just replace the $i \Gamma_{1...7}$ 
by $\Gamma_{1...4}$ for a 4-brane along $X^{5..8}$ hyperplane.

The relative as well as overall signs in eqn. (\ref{9d1pq})
as well as others below 
can be fixed by comparing with 10-dimensional results. 
We also like to mention that all these relations are 
ambiguous upto the choice of spinor basis and do not
affect the counting of the number of supersymmetries in 
network constructions. 

Since the supersymmetry conditions for strings and 4-branes
define a map from physical space to the internal space, 
one can form networks with these objects having
arbitrary charges in a similar way as in ten dimensions. 

\subsubsxn{Membranes and Three branes}

Membranes and U3-branes couple electrically and
magnetically to the 3-form field in nine dimensions which is
a singlet under this $SL(2)$.
As a result, these branes are also invariant under $SL(2)$. 
In fact, they can be obtained from the
ten dimensional IIB theory by considering the D3-brane
either along $S^1$ or perpendicular to it. From
M-theory side, the membrane is the M2-brane and 
U3-branes can be obtained by wrapping M5-branes on $T^2$. 
The invariant supersymmetry for membranes along {(78)} plane
can be written as:
\be
                    \epsilon_\pm = -i \Gamma_{1..6}
                        \epsilon_\pm.
                                \label{9d2}
\ee
The condition for the U3-brane is also similar. 
For a 3-brane along (678) direction
the condition on the supersymmetry parameter is
\be     
                \epsilon_\pm =  \Gamma_{1..5}
                        (\epsilon_\pm).
                                \label{9d3}
\ee
As an application, these branes can be attached at the ends
of the external branes of a network and will lead to BPS states
in the corresponding gauge theory\cite{berg}.

\subsubsxn{Particles and Five branes}

The particles and the U5-branes couple to 1-form fields.
As mentioned earlier, there are three of them which 
form a doublet as well as a singlet. 
One obtains $SL(2)$ doublets of branes 
by wrapping $(p,q)$-strings on $S^1$
or reducing $(p,q)$ 5-branes in a direction orthogonal 
to $S^1$. The singlet particle is the KK momentum mode
and the U5-brane is its magnetic dual.
In M-theory side, doublet of particles 
are the KK momentum modes arising due to compactification
on $T^2$ while the singlet is the wrapped M2-brane. Also, 
doublet of U5-branes are the KK magnetic monopoles 
while the singlet is simply the M5-brane when reduced to 
nine dimensions. 

We can now write down
the conditions on supersymmetry parameters for the doublet. 
These are given by,
\bea
 \left(\begin{array}{c} \epsilon_+ \\
            \epsilon_- \end{array}\right)
        &=& i\2mat{0}{e^{i \alpha}}{e^{-i \alpha}}{0}
        \Gamma_{1..8} 
        \left(\begin{array}{c} \epsilon_+ \\
            \epsilon_- \end{array}\right)
      \quad\quad {\mathrm and} 
\nonumber   \\
 \left(\begin{array}{c} \epsilon_+ \\
            \epsilon_- \end{array}\right)
        &=& \2mat{0}{e^{i \alpha}}{e^{-i \alpha}}{0}
        \Gamma_{1..3} 
        \left(\begin{array}{c} \epsilon_+ \\
            \epsilon_- \end{array}\right)
\eea
for U-particles and U5-branes respectively. 
Similarly, the singlets satisfy the supersymmetry conditions:
\be
\epsilon_\pm = \Gamma_{1...8} \epsilon_\pm
\quad\quad {\mathrm and} \quad\quad 
\epsilon_\pm = -i\Gamma_{1...3} \epsilon_\pm. 
\ee
These conditions individually 
lead to one-half BPS states. However a particle
(or 5-brane) charged with respect to both, doublet and singlet 
fields, corresponds to one-quarter BPS states.

We have therefore discussed different U-branes in nine dimensions
and their supersymmetries. As
we now go down to lower dimensions, we 
encounter larger $U$-duality groups and networks
with more generalized structure.


\subsxn{Eight Dimension}


We now discuss the situation in eight-dimensional
space-time. The $U$-duality symmetry in the present
case is given by $SL(3)\times SL(2)$. The massless 
spectrum contains a set of 5 scalars which form the
$SL(3)/SO(3)$ moduli and a set of two scalars corresponding 
to $SL(2)/U(1)$ moduli. 
Apart from these, one has 1-form and 2-form fields
in (3,2) and (3,1) representations respectively.  They couple 
to particles (or U4-branes) and strings (or U3-branes)
respectively.

The scalars can be arranged in a vielbein form. 
The $SL(2)/U(1)$ part of the vielbein is same as in 
ten dimensions, while the
SL(3) part can be written as an upper triangular matrix 
$\lambda_i^{~a}$ with unit determinant 
and ($i$, $a$) are $SL(3)$ and $SO(3)$ indices respectively.
This is given by
\be
\lambda_i^{~a} = \3mat{e^{-{(\phi + \alpha)\over 2}}}
{\chi e^{\alpha \over 2}}{e^{\phi \over 2}
\eta_1 }{0}{e^{\alpha \over 2}}{e^{\phi \over 2}
\eta_2 }{0}{0}{e^{\phi \over 2}}   ,
\ee
where $\phi$, $\alpha$, $\chi$, $\eta_1$, $\eta_2$
are the five scalar fields parametrizing the moduli.
Two $SL(2)$ subgroups correspond to setting $\phi = \eta_1 = 
\eta_2 =0 $ and $\phi = -\alpha$, $\chi = \eta_1 = 0$. 
Under a generic $SL(3)$ transformation
$\Lambda$ the transformation of the scalar fields 
can be given by
\be
\lambda_i^{~a}(\mu) \longrightarrow 
\lambda_i^{~a}(\mu^\prime) =
\Lambda_i^{~j}\lambda_j^{~b} \tilde{\cal O}_b^{~a}
\ee
where, $\mu$ are the scalar parameters and ${\cal O}$
represents the $SO(3)$ matrix corresponding to the $SL(3)$
element $\Lambda$. 


In eight dimensions, the $N=2$ supergravity is
parameterized by four pseudo-Majorana spinor parameters
transforming in $(2_+, 2_-)$ representation of the maximal
compact subgroup of $U$-duality 
group which is $SU(2)\times U(1)$.
Explicitly, they are given by:
$\epsilon^1_{\pm}$ and $\epsilon^2_{\pm}$ 
where $\pm$ are the $U(1)$ charges. Under 
a generic $SL(3)$ transformation $\Lambda$,
they transform as :  $\epsilon_+ \rightarrow U
\epsilon_+$ and  
$\epsilon_- \rightarrow U \epsilon_-$, where the 
spinors $\epsilon_{\pm}$ are defined as:
\be
        \epsilon_{\pm} \equiv 
                \pmatrix{\epsilon_{\pm}^1 \cr
                        \epsilon_{\pm}^2},
                        \label{defeps}
\ee
and the transformation $U\in SU(2)$ can be obtained from 
the vielbein. These supersymmetry parameters in eight 
dimensions are obtained by the compactification of the 
10-dimensional spinors $\epsilon_{L, R}$. We now present the 
supersymmetry conditions for various U-branes in 8-dimensions. 
 
\subsubsxn{Strings and Three Branes}

In eight-dimensions, U-strings  
couple electrically to
three antisymmetric tensor fields which 
tranform in a $(3, 1)$ representation under 
$SL(3)\times SL(2)$.
Each string is therefore denoted by three $SL(3)$ 
charges $(p, q, r)$. {From} the M-theory point of view, 
this can be thought of as the M2-brane wrapped on
the $(p, q, r)$ one-cycle of $T^3$. The type 
IIB $SL(2)$ is contained in 
$SL(3)$ and the extra $SL(2)$ corresponds to 
the constant deformations of $T^2$ on which
the type IIB is compactified. So from the type 
IIB point of view these strings correspond 
to the ten dimensional $(p, q)$ strings 
and the wrapped 3-branes .

A detailed analysis of the supersymmetry 
property of the string case has been done in 
a recent paper by the present authors\cite{bhatt}. 
The supersymmetry parameters for a string 
along $X^7$ satisfy: 
\be     
\pmatrix{\epsilon_+\cr \epsilon_-} = 
        - i \pmatrix{ {\chi \over |X|} & 0 \cr 
                0 & {\chi \over |X|}}
\Gamma_{1..6}           \pmatrix{\epsilon_+\cr \epsilon_-},
                        \label{8d1} \ee
where $\chi$ is a $2\times 2$ matrix defined in terms of an 
$SO(3)$ vector as: $\chi = X_i \sigma_i$. The components of
the charge vector $\vec{X}$ 
are given in terms of $SL(3)$ charges $p_i$'s and 
$SL(3)/SO(3)$ moduli $\lambda_{i}^{~ a}$ as 
$X_i = \lambda^{-1}_{i a} p_a$. $|X|$ 
is the magnitude of this vector. Above condition is
then manifestly $SL(3)$ invariant.

One can also obtain this relation in a similar 
way as the $(p, q)$ string. This is done by starting from the 
supersymmetry condition for $(1, 0, 0)$ with
trivial moduli 
and then considering an $SO(3) \in SL(3)$ transformation 
that will generate $(p,q,r)$
charges but keep the moduli fixed. Finally, 
to have non-trivial moduli, one 
makes another $SL(3)$ transformation given 
by the vielbein $\lambda_i^a$.

The objects hodge dual to strings in 8-dimensions
are the U3-branes. From the M-theory 
point of view, they can be seen
as M5-branes wrapped on $(p, q, r)$ 2-cycles. From type IIB side
these are the combinations of the wrapped 
$(p,q)$ 5-branes on $T^2$  and the
3-brane. The U-duality invariant supersymmetry
condition for a 3-brane
is similar to the one in eqn.(\ref{8d1}). 
The only modification is the 
number of $\Gamma$ projections. For a 
3-brane along $X^{567}$ hyperplane 
we therefore have:
\be
        \pmatrix{\epsilon_+\cr \epsilon_-} = 
         \pmatrix{ {\chi \over |X|} & 0 \cr
                0 & -{\chi \over |X|}} \Gamma_{1..4}
                \pmatrix{\epsilon_+\cr \epsilon_-}.
                        \label{8d3}
\ee

These invariant relations lead to the fact that one can
place  many strings or 3-branes with arbitrary charges in a manner 
keeping some supersymmetry unbroken,
provided they are properly oriented.
All these states preserve one-eighth or more supersymmetry.

\subsubsxn{Four Brane and Zero Brane}
 
The objects that couple to the 1-forms are the 
0-branes and 4-branes. From M-Theory point of view
they are the M2-branes (M5-branes)
wrapped on different two-cycles (one-cycles)
together with the KK momentum modes (its magnetic dual) 
along three $S^1$. From type IIB side 
they are $(p,q)$ strings (and 5-branes) with various
wrapping and the KK momentum modes. They transform in the $(3,2)$
representation of the $U$-duality symmetry. So the charges
can be writen as $Q_{ia}$ where $i$ and $a$ are the SL(3)
and SL(2) vector indices respectively. 

The invariant supersymmetry can now be 
written by observing the form of equations (\ref{8d1}) and
(\ref{10d1pq}). In terms of $SL(3)\times SL(2)$ charges and 
moduli, the supersymmetry condition for a 0-brane is:
\be
        \pmatrix{\epsilon_+\cr \epsilon_-} = 
        {1\over\Delta}\2mat{0}{-Q}{\bar Q}{0}
        \Gamma_{1...7}
        \pmatrix{\epsilon_+\cr \epsilon_-},  
\label{8d0}
\ee
where we have written the $SL(3)\times SL(2)$ charges
in the form of complex $2\times2$ matrices
$Q = Q_i\sigma_i=(Q_{i1}+iQ_{i2})\sigma_i$.
For the special case 
$n_i=i\epsilon_{ijk}Q_j\bar Q_k =0$
the phase factors of the $Q_i$'s
are equal. Eqn. (\ref{8d0}) then implies the 
breaking of one-half supersymmetry.
However, for $n_i\ne 0$ one has to impose an
additional condition namely,
$n_i \sigma_i \epsilon_{\pm} = \pm |n| \epsilon_{\pm}$ 
and therefore only one-quarter supersymmetry is preserved. 
In eqn. (\ref{8d0}) $\Delta$ is a normalization factor
given by $\Delta^2 = Q_i\bar Q_i + |n|$.

For one-half BPS states the supersymmetry
condition can be written in a simpler form 
as
\be
        \pmatrix{\epsilon_+\cr \epsilon_-} = 
         \pmatrix{ {\chi \over |X|} & 0 \cr
                0 & {\chi \over |X|}} 
                \pmatrix{0 & -e^{i \alpha}\cr
                e^{- i \alpha} & 0 }
                \Gamma_{1..7}
                \pmatrix{\epsilon_+\cr \epsilon_-},
                        \label{8d02}
\ee
where the phase $\alpha$ is defined in terms of  
$SL(2)$ charges and moduli as before and 
$\chi = Q e^{- i \alpha}$. 

For the 4-branes, the invariant 
supersymmetry condition can be
written just by replacing  $\Gamma_{1..7}$ 
by  $i\Gamma_{1..3}$ and $-Q$ by $Q$ keeping $\tilde Q$ 
unchanged in eqn. (\ref{8d0}). Since these are again 
extended objects, one can construct BPS
webs using properly charged 4-branes.

\subsubsxn{Membranes}

Membranes or U2-branes are the dyonic objets 
in eight dimensions and couple to the 3-form fields. 
In type IIB side these are 3-branes wrapped
on a $(p,q)$ cycle of $T^2$. Since the 3-form
field in eight dimensions comes from the self-dual
4-form in ten dimensions, membranes form an $SL(2,Z)$
multiplet. These membranes can also be thought of as coming
from the M2-brane as well as wrapped M5-brane
and is invariant under the $SL(3)$ part of
the $U$-duality group. The supersymmetry condition
can then be written as
\be
        \pmatrix{\epsilon_+\cr \epsilon_-} = 
         \pmatrix{0 & e^{i \alpha}\cr
                e^{- i \alpha} & 0 }
                \Gamma_{1..5}
                \pmatrix{\epsilon_+\cr \epsilon_-}  ,
                        \label{8d2}
\ee
where the parameter $\alpha$ is defined as 
$\alpha=\arg(p+q\tau)$ like the case of string in ten dimension.
Since these membranes form an $SL(2)$ multiplet, one can construct
the network of membranes preserving one-quarter supersymmetry.

We have therefore given the  supersymmetry 
conditions for various U-branes in eight dimensions. 
As mentioned, one can reproduce the
10-dimensional results for both type IIA and IIB string theory. 
In other words the branes in lower dimensions combine the 
results for the type IIA as well as IIB theory. Moreover, the 
8-dimensional branes in the presence of moduli and arbirary 
charges have non-perturbative information about 
string theory, which is not contained in the original 
10-dimensional branes. 


\subsxn{Seven dimension}


After discussing the 8-dimensional case, we now extend these
results to other lower dimensional cases. In 7-dimensions, 
the $U$-duality symmetry is $SL(5)$. 
The massless spectrum contains 
14 scalars, ten 1-form fields and five 2-form fields.
So we have all the branes starting from (-1)-brane (instanton)
to 4-brane. Under $SL(5)$ the ten 1-form fields 
transform in an antisymmetric representation of 
$SL(5)$ and the five 2-form fields transform as 
vector. The 14 scalars, transform 
under both the $SL(5)$ and its maximal 
compact subgroup $SO(5)$ as the $SL(5)/SO(5)$ coset.
To get their explicit transformation we write down the $5\times 5$
upper triangular matrix with unit determinant: $V_i^a$ is 
parametrized by the 14 scalars which represent the vielbein 
for $SL(5)/SO(5)$.
The transformation is given by:
\be
V_i^{~a}(\mu) \longrightarrow 
V_i^{~a}(\mu^\prime) = g^{-1}_{ij}
V_j^{~b}(\mu)\tilde{\cal O}_{ba},
\label{7dmod}
\ee
where as before $\mu$ and $\mu^\prime$ are the moduli 
parameter and indices $i$ and $a$ correspond to $SL(5)$ 
and $SO(5)$ respectively. 

The supersymmetry parameters correspond to 4 spinors which
transform among themselves in the 4-dimensional spinor 
representation of $SO(5)$.
\be
\Lambda : \epsilon^a \rightarrow U^a_{~b}\epsilon^b, 
\label{7dsusy}
\ee
Where $U$ is an $SO(5)$ element in the spinor representation.
We now analyse the supersymmetry properties of 7-dimensional branes.

\subsubsxn{Particle and Three Brane}

The 0-brane (and 3- brane) couple to ten vector fields 
which transform in a 10-dimensional representation of the 
U-duality symmetry $SL(5)$. These can be thought of as 
the M2-brane (and M5-branes) wrapped on 
different two-cycles of $T^4$ together with the  
KK momentum modes arising due to the compactification.
In type IIB side they are wrapped strings (5-branes)
on one-cycles (two cycles) and the corresponding KK momentum
modes.

The charge of the 0-brane is given by a $5\times 5$ antisymmetric
tensor $Q_{ab}$ which transforms as 10 of $SL(5)$. To write down 
the supersymmety 
condition in the present case we define a tensor $\chi_{i j}$
as 
\be
        \chi_{i j} = \lambda^{-1}_{i a} \lambda^{-1}_{j b}
                        Q_{a b}
                                \label{tensor}
\ee
where $\lambda_{i a}$ are the vielbein representation of the 
$SL(5)/SO(5)$ moduli.

Now, to write 
down the supersymmetry condition, we define a matrix 
$\hat{\chi}$ in terms of $SO(5)$ Gamma matrices $\gamma^i$'s 
as: 
\be
        \hat{\chi} = \chi_{i j} \gamma^{i j}.
                                \label{hatchi}
\ee

These constructions are the generalizations of 
similar ones for $SL(3)$ vectors
in eight dimensions. The $SL(5)$ invariant supersymmetry condition 
is then written as:
\be
        \epsilon = {\hat{\chi}\over \Delta} 
        \Gamma_{1..3} \epsilon,
                                \label{7d3}
\ee
where $\Delta$ is a normalization factor obtained the same 
manner as in the case of 0-branes in eight dimensions. 
In other words, the tensorial nature of $\gamma^{i j}$ is 
compensated by that of $\chi_{i j}$ to give an $SL(5)$ invariant. 
The amount of the supersymmetry broken will depend
on the charge. If 
$n_m\equiv\epsilon_{ijklm}\hat \chi_{ij}\hat \chi_{kl} = 0$ 
then it breaks one-half supersymmetry whereas 
$n_m\neq 0$ will preserve only one-quarter. 
This condition is derived from the consistency of 
eqn.(\ref{7d3}) under multiple operations and is similar to the one
in \cite{malda}, where it was obtained from a general classification
of branes using properties of central charges.

The condition for one-half BPS states (\ref{7d3}) can be generated 
from that for a 3-brane in ten dimesions by applying 
$SL(5)$ transformations
as in the earlier cases. 
The U3-branes magnetically couple to 1-form 
fields in seven dimensions and 
satisfy the same supersymmetry condition 
(\ref{7d3}) with $\Gamma_{1..6}$ 
replaced by $\Gamma_{1....3}$.

\subsubsxn{String and Membrane}

In seven dimesions one also has U-strings and 
membrane which are dual to
each other and have charges which transform as a vector under
SL(5). The M2-brane reduced to seven dimensions and the
M5-branes wrapped on four 3-cycles constitute the multiplet 
of U2-brane. In type IIB side these are 3-branes 
on three 1-cycles and $(p, q)$ 5-brane on a 3-cycle. 
Similarly the U-string is coming from 
M2-brane on four 1-cycles and M5-brane
on one 4-cycles while in type IIB it is a $(p, q)$ string 
alongwith the D3-brane on three 2-cycles. 
U2 and U1-branes in seven dimensions satisfy similar conditions.
Their supersymmetry property is similar to 
that of a U-string in 8-dimensions since both
have charges which transform as a vector.

Since the supersymmetry parameters transform in a
spinorial representation while the charges are in vector one, 
to define the
invariant supersymmetry condition we introduce 
$SO(5)$ charges in a spinor representation as a 
matrix: $\phi \equiv Y_i \gamma_i$, 
where $Y_i = V_i^{~a}Y_a$. $V_i^a$'s are the $SL(5)/SO(5)$
vielbeins and $Y_a$'s are the $SL(5)$ charges. Now one can  
write down the
invariant supersymmetry relations for strings and U2-branes as:
  \be
        \epsilon = {\phi\over {|\phi|}}\Gamma_{1..5} \epsilon
\quad\quad\quad
{\mathrm and}\quad\quad\quad 
        \epsilon = {\phi \over {|\phi|}}\Gamma_{1..4} \epsilon
                                \label{7d12}
\ee  
repectively. These supersymemtric conditions are also a simple 
generalizations of the result in \cite{bhatt} and in 
eqn. (\ref{8d1}). In addition, in seven dimensions, 
there are 14 scalars parameterizing the $SL(5)/SO(5)$
which couple to the instantons and
4-branes. Their explicit solution and supersymmetry condition
will also be of interest to analyse.

We have therefore dicussed different $U$-branes in seven 
dimensions. In this case, we have also seen the application of 
tensorial charges in obtaining the invariant 
supersymmetry condition. The charges span a five
dimensional internal space and such  branes can therefore
give ries to five dimensional webs with 1/32 or more supersymmetry.


\subsxn{Six Dimension}

We now discuss the case in six dimensions. The $U$-duality 
group
is $SO(5, 5)$. The massless spectrum contains 
25 scalars parametrizing 
the coset $SO(5,5)/SO(5)\times SO(5)$, sixteen
1-form fields ($A_{\mu}$) in a 16-dimensional spinorial representation 
of $SO(5,5)$ and five 2-form fields $(B_{\mu\nu})$ 
forming a 10-dimensional representation of
$SO(5, 5)$ by decomposing the $B_{\mu \nu}$'s into self-dual 
and anti-self-dual forms. 
The six dimensional theory therefore 
has the brane configurations: U3-brane and U-instantons
which couple to the scalars, U2 and U0-branes which couple to vectors
and finally U1-brane coupling to the ten 2-form fields. 

Now consider the representations of different fields. 
The maximal compact subgroup of $SO(5,5)$ is $SO(5)\times SO(5)$.
The scalars, as in earlier cases, transform under both $SO(5,5)$ 
and $SO(5)\times SO(5)$. They can be written down 
in the  form of a vielbein
given by \cite{tani}:
\be
U = \2mat{a_{\hat m}^a}{b_{\hat m}^{\dot a}}{c_{\dot m}^a}
{d_{\dot m}^{\dot a}} \label{spinor},
\ee
where 
$a_{\hat m}^a$, ${b_{\hat m}^{\dot a}}$, ${c_{\dot m}^a}$, 
${d_{\dot m}^{\dot a}}$ are $5\times 5$ matrices, 
$\hat m , \dot m (= 1 ..5)$ together are the $SO(5,5)$ vector indices
and $a$ and $\dot a$ are the vector indices for the 
two $SO(5)$'s respectively. The `dot' is to distinguish 
the second $SO(5)$  from the first. 
The vielbeins transform as:
\be
U(\mu) \longrightarrow U(\mu^\prime) =
g U(\mu) h^{-1}, \quad {\mathrm where} 
\quad g\in SO(5,5), ~~~ h\in SO(5)\times SO(5)
~~
\ee
and $\mu$ is the scalar parameters.
As in other cases, using these moduli 
one can associate an $SO(5,5)$
element with an $SO(5)\times SO(5)$ one.

Since the vector fields 
transform in the Majorana-Weyl spinor 
representation of $SO(5,5)$ we present it explicitly.
To define those representation let us consider the 
Clifford algebra for $SO(5,5)$.
The $SO(5,5)$ gamma matrices are given by tensor product 
of two $SO(5)$ ganmma matrices and pauli matrix as
\bea
\Gamma^{\hat m} &=& \gamma^{\hat m}\times 
1\times\sigma_1  ,\quad\quad  \hat m = \hat 1,\cdots,\hat 5, 
\quad\quad {\mathrm and}\nonumber  \\
\Gamma^{\dot m} &=& 1\times \gamma^{\dot m}
\times -i\sigma_2  , \quad\quad  \dot m = \dot 1,\cdots,\dot 5.   
\eea
We define $\Omega$ as $\gamma^T = 
\Omega \gamma \Omega^{-1}$, where $\gamma$ is a 
SO(5) Gamma matrix. The $\Omega$ serves as the 
raising and lowering operator for SO(5) spinor 
index : $\psi^\alpha=\Omega^{\alpha\beta}\psi_\beta$.
The SO(5,5) Gamma matrices
satisfy
\bea
\Gamma^{m*} &= 
B\Gamma^m B^{-1} , \quad {\mathrm where}\quad
B = \Omega\times\Omega\times 1 , \quad {\mathrm and} ,\\
\Gamma^{mT} &= 
C\Gamma^m C^{-1} , \quad {\mathrm where}\quad
C = \Omega\times\Omega\times \sigma_1 .
\eea
A 32 component spinor $\psi$ of $SO(5, 5)$ consists of
two
$SO(5)\times SO(5)$ spinors $\xi_{\mu\dot\mu}$ and 
$\chi_{\mu\dot\mu}$ which are of opposite chirality .
If we impose the Majorana-Weyl condition:
\be
\bar\Gamma\psi \equiv -\psi ,\quad\quad  
\psi^* = B\psi  , \quad\quad {\mathrm where} 
\quad\quad
\bar\Gamma = \Gamma^{\hat 1}\cdots\Gamma^{\hat 5}
\Gamma^{\dot 1}\cdots\Gamma^{\dot 5} 
\ee
we get $\psi = \left({0\atop\chi}\right)$. Under $U$-duality group it
tarnsforms as
\be
\chi_{\mu\dot\mu} \longrightarrow
(S^T)_{\mu\dot\mu}^{~\nu\dot\nu}\chi_{\nu\dot\nu}
\ee
where $S_{\mu\dot\mu}^{~\nu\dot\nu}$ is the
spinor tranformation matrix of $SO(5)\times SO(5)$.

In order to write down the invariant supersymmetric relation
we have to transform the $SO(5,5)$ charges into 
$SO(5)\times SO(5)$  charges. This can be achieved 
by using an 
alternative form of moduli which is a  
$16\times 16$ matrix denoted by $V_{\mu\dot\mu}
^{\alpha\dot\alpha}$, where $\alpha$ and 
$\dot\alpha$ are the $SO(5)\times SO(5)$ 
spinor indices. These moduli are related to the 
vector moduli in the following way \cite{tani}:
\be
a_{\dot m}^a = {1\over 16}V^{\mu \dot\mu \alpha \dot\alpha}
(\gamma_{\hat m})_\mu^\nu(\gamma^a)_\alpha^\beta
V_{\nu \dot\mu \beta \dot\alpha} \quad\quad {\mathrm etc.}
\ee

The supersymmetry parameters transform only 
under the maximal compact
subgroup $SO(5)\times SO(5)$ . One can write them in two
sets of four Majorana-Weyl spinors in six dimension 
$\epsilon^1$ and $\epsilon^2$ which
transform in a representation $(4,1)$ and $(1,4)$. Using 
the vielbeins
one can associate a unique $SO(5)\times SO(5)$ element 
with a generic $SO(5,5)$ which determines the 
transformations like that in other dimensions. 
We now discuss the supersymmetry conditions for
U-branes in six dimensions.

\subsubsxn{Particle and Membrane}

Let us consider the particles and membranes which are dual to each other
in six dimensions. They have the 16 charges in the spinor
representation of $SO(5,5)$. Sixteen U2-branes from M theory point of view
are coming from the M2-brane and wrapped M5-brane on ten
3-cycles of $T^5$ and five KK momentum modes. From 
type IIB side they are 
$(p,q)$ 5-branes wrapped on four 3-cycles of $T^4$, 3-branes on four 
1-cycles and four KK momentum modes. The U0-branes come from dual 
configurations. In order to write down a supersymmetry 
configuration we need
to transform the $SO(5, 5)$ spinor charges of the vector fields into the
spinors of  $SO(5)\times SO(5)$. 

Making use the moduli $V_{\mu \dot{\mu}}^{\alpha \dot{\alpha}}$,
one can define $SO(5)\times SO(5)$ spinor charges 
$Q^{\alpha \dot{\alpha}}$ from the original $SO(5, 5)$ charges 
$Q_{\mu \dot{\mu}}$. One can then write a U-duality invariant 
supersymmetry condition for U2-branes in six dimensions as:

\be
        \epsilon_\alpha  = {1\over\Delta}Q_{\alpha \dot\alpha} 
        \Gamma_{1..3}\epsilon^{\dot \alpha} 
\quad\quad {\mathrm and}  \quad\quad 
        \epsilon^{\dot\alpha}  =-{1\over\Delta}\tilde Q^{\dot\alpha\alpha} 
        \Gamma_{1..3}\epsilon_\alpha    
                                \label{6d2}.
\ee
where $\tilde{Q}^{\alpha \dot{\alpha}}$ is simply the contravariant
tensor corresponding to $Q_{\alpha \dot{\alpha}}$ and indices are
raised and lowered using $\Omega$'s defined earlier. 
$\Delta$ is again a normalization factor.
Once again a similar condition will be satisfied by a U0-brane, 
with Gamma matrix projection being: $\Gamma_{1..5}$
and $-\tilde Q$ is replaced by $\tilde Q$. 

The consistency between two equations in (\ref{6d2})
requires $Q \tilde Q = \tilde Q Q = 1$. One can argue that 
when charges are restricted in this manner, 
the supersymmetry condition corresponds to one-half BPS state.
Otherwise, supersymmetry will be broken further. 
Once again similar results appear in other contexts 
as well\cite{malda}.

\subsubsxn{String}

Finally, we discuss the situation with strings 
(U1-branes) in six dimensions.
This case is similar to the one in four dimensions, where
one has both the electric and magnetic charged objects. 
Here, one can 
combine the "electric" and "magnetic" field strengths 
and form a ten dimensional vector under $SO(5, 5)$.

The charges, which trasform as vectors $(5, 1) + (1, 5)$ 
under the $SO(5)\times SO(5)$ symmetry group are 
constructed by using the
vielbeins $a_{\hat m}^a$, $b_{\hat m}^{\dot a}$ etc. 
in eqn.(\ref{spinor}). We denote these 
$SO(5)$ vectors as $X^1_a$ and $X^2_{\dot a}$ 
respectively. The invariant supersymmetry conditions are 
then similar to the ones in higher dimension for a string 
and can be written as:
\be
        \left(\begin{array}{c}\epsilon_1 \\
        \epsilon_2 \end{array} \right)
        = \2mat{\chi^1}{0}{0}{\chi^2}
                        \Gamma_{1..4} 
        \left(\begin{array}{c} \epsilon_1 \\
         \epsilon_2 \end{array}\right). 
                                \label{6d1}
\ee
where, $\chi^1 = X^a\gamma_a$ and 
$\chi^2 = X^{\dot a}\gamma_{\dot a}$.
If the SO(5,5) charge vector $(X^{\hat m}, X^{\dot m})$ is null,
$\chi^1$ and $\chi^2$ can be simultaneously
normalized to unity. Then this condition breaks one-half
supersymmetry. Otherwise supersymmetry breaks to one-quarter, 
as the consistency of eqn. (\ref{6d1}) then requires setting 
all the components of $\epsilon_1$ or $\epsilon_2$ to zero.


\subsxn{U-branes in Four and Five Dimensions}

In five dimensions,
the full U-duality symmetry is a non-compact
version of $E_6$. The massless $p$-form fields consist of 
42 scalars and 27 vector fields. So it has the U(-1)-branes 
and strings as well as their duals namely, U2 and U0-branes. 
They couple to the scalar and vector fields respectively..

In this case the maximal compact subgroup is
$USp(8)$. The 42 scalars can be written as a vielbein
$V_{\alpha\beta}^{~~ab}$ for the coset space $E_6/USp(8)$,
where $\alpha$, $\beta$ are the $E_6$ indices and $a$, $b$
are $USp(8)$ indices. The transformation is once again like
\be 
V(\mu) \longrightarrow V(\mu^\prime) = g V(\mu) h^{-1} , 
\quad {\mathrm where} \quad g\in
E_6 , \quad h \in USp(8).
\ee
The 27 vector fields $A_\mu^{\alpha\beta}$ transform in the 
fundamental representation of $E_6$ and are singlets of the 
$USp(8)$.  The supersymmetry parameters are given by 
8 spinors in five dimensions. Under $USp(8)$ 
they transform in the fundamental representation.

The U-string and the U0-brane in this 
case couple to the 1-form field
and so the charges transform in the fundamental of $E_6$.
To find out the invariant supersymmetry, 
one uses the moduli $V_{\alpha \beta}^{a b}$, 
to transform the charges into a 27-dimensional 
representation of $USp(8)$.  We denote these charges as:
$X_{a b}$, with raising and lowering defined by the 
symplectic metric $\omega^{ab}$. 
The simplest case now preserves one-half supersymmetry
and the invariant condition for U1-brane has a form:
\be
        \epsilon^a = {X^{a b}} 
        \Gamma_{1..3} \epsilon_b,
                        \label{5d1}
\ee
where $X_{a b}$ are the $8\times8$ antisymmetric matrices
satisfying $\omega^{a b} X_{a b} = 0$ and are properly 
normalized. 
The U0-brane can be found by replacing $\Gamma_{1..3}$
by $i\Gamma_{1..4}$ in (\ref{5d1}). In the $1/2$ supersymmetry case it simply 
corresponds to $X^2 = -1$.
It is expected that there are other
charges as well which breaks supersymmetry to 
one quarter and one eighth. However we postpone these discussions 
for future.



By continuing to 4-dimensions, the U-duality group
is $E_7$ with maximal compact subgroup: $U(8)$.
Here we have 28 vector fields and 70 scalar fields.
One can form the $E_7/U(8)$ vielbeins using these scalar fields.
The vielbeins are 56 dimensional matrices which transform under 
$E_7$ as 56 while under $U(8)$ it transforms as $28 + \bar{28}$.
The vector fields
are the self-dual fields in 4-dimensions. To get the transformations
one has to split the 28 vector field strengths in self-dual and
anti-self-dual parts. These electric
and magnetic fields then combine 
to give a 56-dimensional representation of $E_7$.
The sixteen supercharges $\epsilon^L$ and $\epsilon^R$ 
tranform as a $8_{1\over 2}$ and
$\bar{8}_{- 1\over 2}$ representation of $U(8)$.  One also has
70 scalar fields which paramaterize a coset $E_7/SU(8)$. 
These scalars can be parameterized by a $56\times 56$ matrix, 
which plays the role of a vielbein in the present situation.
In particular, it transforms as a 56-representation of $E_7$
and $28 + \bar{28}$ representaion of $SU(8)$. Using these 
vielbein's one can construct the charges
with tensor structure $28$ and $\bar 28$ of $SU(8)$.
We represent these charges as
$\chi_{ij}$ and $\chi^{ij}$. 
It is then straightforward to write down the supersymmetry conditions
for the dyonic 0-brane in four dimensions. 
For a U0-brane, we have:
\be
       \epsilon^{L}_i = \chi_{ij} \Gamma_{1..3}
	\epsilon^{R~j} ,
\quad\quad
       \epsilon^{R~i} =- \chi^{ij} \Gamma_{1..3}
	\epsilon^{L}_j , 
                                \label{4d0}
\ee
and a similar condition for $\epsilon^{R}$.
This condition represents the one half BPS states
when $\chi^{ij}\chi_{jk} = -\delta^i_k$.
However generalizations to cases
preserving less amount of supersymmetries should also be 
possible. 

We now apply the results derived in this section 
for the construction of network and web of U-branes.

\sxn{Applications}

So far we have mainly discussed supersymmetries of 
individual $U$-branes with generic charges in various dimensions.
We have seen that each of them preserves some amount of 
supersymmetry.  
It will be interesting to obtain such U-brane solutions
directly from supergravity and verify our results.
More interesting objects are the intersecting
$U$-branes, as well as the network or web type configurations.  
By using the supersymmetry conditions
one can construct various intersecting brane configurations
and webs which represent BPS states. It should also be  
possible to construct the corresponding supergravity solutions.

We now discuss the applications of the results in the previous 
section to construct general network like configurations in 
various dimensions. The requirements of such constructions is
the presence of a certain amount of supersymmetry when an 
arbitrary number of such branes are put together. 
In this connection, interesting objects are the 
branes with $U$-duality charges. Then one can form 
junction where three branes meet. In general they can meet 
along any hypersurface with dimension smaller than the particular
branes involved. One can then  
make use of these junctions to build web structures. Apart
from the supersymmetry condition, each brane-junction has
to satisfy the charge balance and the tension 
balance conditions as well. One can also  
construct junctions of more than three branes at a time,
but these junctions can 
be thought of as a degenaration of a net due to the vanishing
of a polygon face. Another set of interesting objects 
in this case are the nets with branes ending on 
other orthogonal branes.  In this context, 
whether a brane can end on another one will be fixed by the 
$U$-duality group. For example, in 8-dimension $(1,0,0)$ 
string can
end on $(0,1,0)$ as they come from $F$ and $D$ strings of ten
dimension. Other cases, related to this one  
through $U$-duality are therefore also compatible.

For the construction of webs, 
the ten dimensional cases have been well studied. Since the strings 
and five branes have $(p,q)$ charges, one can construct the 
planar webs of strings and five branes. Note that it is not
 possible to
consider a non-planar BPS configuration as the charge space 
is two dimensional. In this case one also has 
D3-branes which are invariant under
$S$-duality group. So one can consider the configurations involving the
3-branes and 5-branes or strings. These configurations gives rise to
interesting gauge theories. Also, one can consider the three 
string junction with strings ending on three D3-branes. These 
objects have been identified as 
the BPS states in $SU(3)$ SYM. 
A gauge theory structure may be possible to write down directly 
for the webs, although they  have not yet 
been constructed explicitly. However zero mode analysis have 
been done \cite{bk,cal} in this context to obtain an S-Matrix
structure for these theories.

We now consider the eight-dimensional case. 
$U$-strings and $U$3-branes now 
have charges which are vector under $SL(3)$. They are neutral with 
respect to the remaining part of the $U$-duality, namely $SL(2)$. 
The $U$4-brane charges
are vector under $SL(3)$ as well as $SL(2)$. 
The $U$2-brane is a singlet of $SL(3)$ and a doublet of $SL(2)$.
The essential idea for constructing webs then is to 
present an identification
between the coordinates of the `internal space' and that of the physical
space. 

For the eight-dimensional case, the non-planar  network of
strings with $1/8$ supersymmetry has been  discussed 
previously\cite{bhatt}.
The internal space is three dimensional and the direction of 
one of the charge vector is given by $\vec{X}$ . The corresponding 
string is then aligned  
along the direction $\hat n$ in the $(123)$ space.
The supersymmetry condition for this general orientation of string
can be written in a covariant form as,
\be
\left( {\epsilon^1_\pm \atop \epsilon^2_\pm } \right)
= -{i\over 2}(\hat n\cdot\sigma)~ e_{ijk} \hat n_i\Gamma_{4..7}\Gamma_{jk}
\left( {\epsilon^1_\pm \atop \epsilon^2_\pm } \right),
                             \label{cov.}
\ee
where $\hat n={\vec X\over |X|}$ and $e_{ijk}$ is
the Levi-Civita tensor. 
Now, any rotation of the string in the 
3-dimensional space, spanned by $(1, 2, 3)$ directions,  
is associated with a rotation in the charge space so as to
preserve the above condition. Then, any number of 
strings can be put together with
appropriate orientation and charges. All of them together
break one-eighth supersymmetry with the condition: 
\be
i\sigma_i\epsilon_\pm = {1\over 2}
e_{ijk}\Gamma_{jk} \epsilon_\pm  ,
\quad {\mathrm and}\quad
\Gamma_{4\cdots 7}\epsilon_\pm = \epsilon_\pm . 
\ee
These networks can also be obtained from 
M-theory compactification on a 
non-trivial geometry  along the line of \cite{leung}.

Similarly, following an earlier work on the web 
structure of 5-branes in ten dimensions, we can also construct webs of 
eight-dimensional U3-branes. For this, one can  take two 
directions of each of the
3-branes as common. The remaining direction can be aligned 
in exactly the same
manner as the strings. This will again give a configuration
with $1/8$ supersymmetry. As the U3-branes intersect along a
two dimensional surface, one should be able to write a 
three dimensional worldvolume theory describing such 
configurations. We however do not understand the full structure
of this gauge theory at the moment. 

We next discuss a web constructed of U4-branes in eight 
dimensions. The basic difference in this case with 
respect to the webs of strings and 3-brane is due to the fact
that $U4$-branes carry both $SL(3)$ and $SL(2)$ charges. 
In this case, to construct a class of supersymmetric configuration, we 
write the supersymmetry  condition for a 4-brane, 
when one of its edges is aligned in a
three dimesnional space, making an angle $\theta$ from 
$x^3$-axis and an angle 
$\phi$ from $x^1$-axis in the $x^1 - x^2$ plane. Similarly, 
we choose another plane spanned by coordinates $x^4-x^5$. 
One more edge of 
the 4-brane lies in this plane making an angle $\alpha$ from the 
$x^4$ axis. The 4-branes are then left only with two common 
directions. The supersymmetry condition of eqn.(\ref{8d02}) then 
has a generalized form: 
\bea
        \pmatrix{\epsilon_+\cr \epsilon_-} = 
        &i& \pmatrix {\hat n\cdot\sigma & 0 \cr
                0 & \hat n\cdot\sigma} 
                \pmatrix{0 & e^{i \alpha}\cr
                e^{- i \alpha} & 0 }
                (\Gamma_1\Gamma_2 cos \theta 
\nonumber \\
               &+& \Gamma_2\Gamma_3 sin \theta cos \phi +
                \Gamma_3 \Gamma_1 sin \theta sin \phi)
                (\Gamma_5 cos \alpha + \Gamma_4 sin \alpha)
                \pmatrix{\epsilon_+\cr \epsilon_-}.
                        \label{8d4}
\eea
               
To obtain projections conditions on the supersymmetry 
parameters for arbitrary angles, we parameterize the matrix 
$\chi$ as in \cite{bhatt}. Then the invariant supersymmetry 
parameters can be shown to satisfy the independent conditions: 
\be
        \Gamma_{12}\epsilon_+ 
        = i\sigma_3\epsilon_+  ,
\quad
        \Gamma_{23} \epsilon_+
        =i \sigma_1\epsilon_+  ,
\quad
        (1+i\Gamma_{45})\epsilon_- = 0  ,
\quad
        \Gamma_5\epsilon_- = \epsilon_+   .
                                                \label{cond2} 
\ee
Together they give rise to solutions preserving 
$1/16$ supersymmetry.
The geometric realization of similar configurations has
been discussed earlier \cite{leung}. 
Once again, it will be interesrting to 
construst the low energy gauge theory for these webs in the line of
$(p, q)$ 5-branes \cite{aha}.

The remaining U-brane in eight dimensions, namely
$U$2-brane, is a singlet under $SL(3)$ part of the 
$U$-duality group. One can therefore consider the networks 
of other U1(U3)-branes where the external branes end on 
parallel or orthogonal $U$2-branes. 
This configuration breaks the supersymmetry by 
further one-half and should correspond to non-trivial
BPS states in the low energy world volume theory of the $U$2-branes.
In eight dimensions we also have U0-branes. They can also
possibly be used for obtaining more exotic  constructions of
 webs. For example,
they can form the vertices in  a web consisting of strings and
 membranes, as
they saturate the  $SL(3)$ charges of a string and $SL(2)$ 
charges of membranes.

We have therefore shown the existence of non-perturbative
configurations consisting of webs of $1, 2, 3$ and $4$-branes.
We again emphasize that these are novel objects which  
do not exist in higher dimensions. 
For example, a U2-brane carries charges with respect to both
$A_{\mu \nu \rho}^+$ and $A_{\mu \nu \rho}^-$. 
Such charged objects do not exist in the ten dimensional theory. 

Let us now consider the situation in seven dimensions. 
In this case, we have $U$2 and $U$3-branes 
with charges transforming as vectors and tensors of $SL(5)$.
The internal space now is five dimensional. By utilizing the full 
five dimensional structure of the charge space one can obtain 
configurations where supersymmetry
is broken to 1/32. However one can restrict oneself to a
subspace of the full five dimensional space and
obtain BPS states with more supersymmetry. The $U$2-branes in seven 
dimensions can be dealt with in the same manner as the strings and 
3-branes in eight dimensions, since their transformation properties
are similar.

The case of U3-branes in seven dimensions is more interestring since
they transform as a higher dimensional representation of the 
U-duality symmetry. We restrict ourselves to the one-half BPS 
U3-branes only. To obtain a web like structure out of them,
we write down the 
supersymmetry condition when a 2-plane of the brane
is oriented in a five dimensional subspace of the 
seven dimensional space-time in special way. By choosing such 
an orientation the supersymmetry condition 
can be written in a form:
\be
        \epsilon = {\hat\chi\over |\chi|} 
                {\chi_{ij}\over |\chi|}\Gamma_1 \Gamma_{ij} \epsilon.
                                \label{7d3w}
\ee
where $\hat{\chi} \equiv \chi_{ij} \gamma^{i j}$ as 
in eqn.(\ref{hatchi}). 
Equation (\ref{7d3w}) is simply a generalization of eqn.(\ref{7d3})
to the case when 3-branes have a special orientation in the five 
dimensional space. The first factor $({\hat{\chi}\over {|{\chi}|}})$
gives the orientation of the 3-brane in the internal space, whereas
a factor ${\chi_{i j}\over |\chi|}\Gamma_1 \Gamma_{i j}$ 
gives its orientation in the spatial directions. The presence of 
the same parameters $\chi_{ij}$ in both places implies the 
alignment of the 3-branes in these directions. In other words,
the parameters $\chi_{i j}$ specify the position of
the 3-branes in the same manner that the angles $\theta$ and $\phi$
parameterized the position of a string in a three dimensional space.

Now, to construct a network structure  
consisting of these branes, we solve the conditions (\ref{7d3w})
by applying projections:
\be
    \Gamma_1 \Gamma_{ij} \epsilon = \gamma^{ij} \epsilon.
                        \label{u38}
\ee
For a generic situation these are ten projection conditions
on $\epsilon$ and therefore break supersymmetry  completely. 
However one can restrict to a subset when some of the moduli 
are set to zero and the 2-plane lies in a smaller subspace. 
One can then have supersymmetric solutions as well.

In six dimensions, to construct a string network, the 
simplest possibility is to choose the $SO(5, 5)$ moduli
such that matrices $\chi_1$ and $\chi_2$ are identified in 
equation (\ref{6d1}). Then the situation becomes identical to 
the one for the string network solution in seven dimensions
with $1/32$ or more supersymmetry. 
However the supersymmetry condition (\ref{6d1}) in six 
dimensions may allow 
a more general possibility.  This is because the left and 
right-moving modes of a string represent independent degrees 
of freedom of a Narain lattice and can 
point in different directions. The Gamma matrices in 
two eqns. (\ref{6d1}) can then be different and may be used for
having a web structure on a compact space. It is however not clear
whether such BPS states are also possible in non-compact
spaces. 
Other possibility in six dimensions is to consider U2-branes.
The charges are now 
in spinor representation  whereas the membrane orientation is 
represented by a vectorial direction. It is again not clear
how to use an arbitrary number of these objects for 
constructing BPS states. However it should be  possible to 
construct them by  switching off certain moduli. 

In five dimensions, the only possible object to be used are 
the strings. The charges now describe a $27$-dimensional vector. 
A simple way to construct the BPS configuration once again will be
to switch off a set of moduli, so that the $27$ of $USp(8)$ breaks into
a representation of orthogonal groups such as 
$SO(6)\times SO(2)$. One can then use 
the standard ways, outlined in several examples 
above, to construct the BPS configurations.

\sxn{Discussion}

To conclude, we have analyzed the supersymmetry properties of  
U-branes in various dimensions. We have shown   
that there exist supersymmetric configurations 
of $U$-branes, their intersections, as well as web structures
carrying different
$U$-duality charges in various dimensions. Each of them represent 
BPS states. In particular, we have presented new BPS 
configurations of string theory presrrving $1/32$ or more 
supersymmetries.  

It should be possible to construct 
some of these solutions
explicitly from supergravity point of view\cite{lu}. 
This is certainly true for the U-branes
and their orthogonal intersections. It seems to be technically  
non-trivial to obtain net or web type solutions from supergravity 
point of view. However, the geometry of the nets is obtainable
explicitly from the consideration of the membranes and five branes 
of M theory \cite{lee}.

It has been suggested that certain stable non-BPS states of 
string theory \cite{bk,nonbps} can be represented as web-like
structures through certain moduli deformations. It will be of 
interest to generalize these for other $U$-branes. 
It will also be interesting to obtain the 
low energy effective theory of these $U$-branes and nets.
In ten dimensions, when 
$\alpha'\rightarrow 0$ the branes decouple from the bulk
supergravity and one can write down a low energy effective theory
which is the SYM (DBI) in respective dimension. It is likely
that a similar limit will exist in other cases as well. 
An approach in this direction  
may be the zero mode analysis of the branes
and nets. This  may lead to new phenomena
in the realm of QFT's which can not be obtained otherwise.

Finally, these $U$-branes and nets can possibly be related to  
compactifications on  certain non-trivial manifolds. $D$5-branes
of 8-dimensions has already been shown to play an essential role
in compactification on $T^3$ fibered Calabi-Yau manifold
\cite{ubrane}. 
Moreover, these $U$-brane junctions were also discussed in 
connection with M-theories on toric geometries.          
There, the network structures are expected to be associated 
with the locus of the vanishing cycles of the
toric hypersurface\cite{leung}. 
In the ten dimensional case
such a relation has already been argued  for 
the 5-brane webs \cite{aha}. The extension of these explicit
mappings to the cases of other networks should be interesting
to study as well.

\sxn{Acknowledgements}

We thank Sandip Bhattacharyya for many useful discussions.

\appendix

\sxn{Appendix}

To obtain the $U$-dulaity invariant conditions on supersymmetry parameters for 
different $U$-branes,
we will start from the conditions
for wrapped M2/M5 branes in eleven dimensions. 

The condition on supersymmetry parameters for
M2-branes (along $X^{9} X^{10}$) and M5-branes (along 
$X^{6}...X^{10}$) in eleven dimensions are :
\be
\epsilon^{(11)} = \Gamma_{1..8}\epsilon^{(11)} ,\quad\quad
\epsilon^{(11)} = \Gamma_{1..5}\epsilon^{(11)} ,
\ee
where $\Gamma$'s and $\epsilon^{(11)}$
are the eleven diemnsional gamma matrices and supersymmetry
parameters respectively. The supersymmetry conditions in lower
dimensions can then be obtained in the following way. 

\begin{itemize}

\item U-string in eight dimensions :

The eleven dimensional Gamma matrices can be written as
\bea
\Gamma_\mu &=& \gamma_\mu\times 1, \quad \mu = 0,1,..,7 
\quad \tilde\gamma = i \gamma_0 ... \gamma_7 \nonumber \\
\Gamma_{7+i} &=& \tilde\gamma\times \sigma_i, \quad i = 1,2,3  .
 \label{Ad81} \eea
Where $\gamma_\mu$ are eight dimensional Gamma matrices
and $\sigma_i$'s are Pauli spin matrices.
A U-string along $X^7$ is an M2 brane along $X^7$
and in some other direction $X^{7+i}$. Using the above
splitting, the condition (\ref{Ad81}) becomes
\be
\epsilon^{(8)}_{\pm} = i\gamma_{1..6}\sigma_i 
\epsilon^{(8)}_{\pm} ,
\ee
where $\epsilon^{(8)}$ are eight dimensional spinors
and $\pm$ are the chirality indices.
The condition (\ref{8d1}) can then be obtained from
(\ref{Ad81}) by making $SL(3)$ rotations mentioned earlier.

\item U0-brane in eight dimensions

A U0-brane can be thought of as a wrapped M2-brane on
$X^{7+i}X^{7+j}$ plane. The corresponding supersymmetry
condition will be
\be
\epsilon_+^{(8)} = \gamma_{1..7}\sigma_i\epsilon_-^{(8)} .
\label{Ad80} \ee
$SL(3)$ rotations will then generalize the $\sigma_i$
in (\ref{Ad80}) to ${\chi \over |X|}$ in (\ref{8d02}). 
Finally one 
also has to make $SL(2)$ 
transformations which lead to the $SL(3)\times 
SL(2)$ invariant relations in (\ref{8d0}).

\item U0-branes in seven dimensions

In this case the eleven dimensional Gamma matrices can be written as
\bea
\Gamma_\mu &=& \gamma_\mu\times \rho_5, \quad \mu = 0,1,..,7 
\quad \rho_5 = \rho_1\cdots\rho_4 \nonumber \\
\Gamma_{6+i} &=& 1\times \rho_i, \quad i = 1,2, 3, 4. 
  \label{Ad70}   \eea
Where $\gamma_\mu$ and $\rho_i$ are Gamma matrices in 
seven and four dimensions respectively.
   
U0-brane in seven dimensions can be obtained by wrapping
M2-brane on some plane $X^{6+i}X^{6+j}$ . The supersymmetry
condition (\ref{Ad70}) then becomes
\be 
\epsilon^{(7)} = \rho_{ij} \gamma_{1..6} \epsilon^{(7)}, 
\ee
where $\epsilon^{(7})$ are seven dimensional spinors.
Moreover, by making $SL(5)$ rotations, $\rho_{ij}$ can be generalized 
to $\chi_{ij}$ in condition (\ref{7d3}), where $\chi^2 = -1$ and 
$\epsilon_{ijklm}X_{ij}X_{kl} =0$. 
This represents $U$-duality invariant supersymmetry
condition for one-half BPS U0-branes in seven dimensions.
It can be generalized further to one-quarter BPS states.

\end{itemize}

\vfil
\eject


\begin{thebibliography}{99}
\newcommand{\np}{{\it Nucl. Phys.} {\bf B}}
\newcommand{\pl}{{\it Phys. Lett.} {\bf B}}
\newcommand{\prd}{{\it Phys. Rev. }{\bf D}}
\newcommand{\prl}{{\it Phys. Rev. Lett.}}
\newcommand{\mpl}{{\it Mod. Phys. Lett. }{\bf A}}
\newcommand{\ijmp}{{\it Int. J. Mod. Phys.}{\bf A}}
\newcommand{\cqg}{\it Class. Quant. Grav.}
\newcommand{\jmp}{\it J. Math Phys.}
\newcommand{\cmp}{\it Comm. Math. Phys.}
\bibitem{sch} J. Schwarz, ``Lectures on superstrings and M 
        theory dualities'', {\it Lectures given at ICTP 
        spring school and TASI summer school}, 
        {\it Nucl. Phys. Proc. Suppl.}, {\bf B55} (1997) 1,
         hep-th/9607201.

\bibitem{mukhi} K. Dasgupta and S. Mukhi, ``BPS nature 
        3-string of junction'', {\pl}{\bf 423} (1998), 261 ,
        hep-th/9711094.

\bibitem{sen} A. Sen, ``String network'', JHEP {\bf 3} (1998), 5 
        hep-th/9711130.

\bibitem{zwib} M. Gaberdiel and B. Zwiebach, ``Exceptional 
        groups from open strings'', hep-th/9709013;
        M. Gaberdiel, T. Hauer and B. Zwiebach, ``Open string-string
        junction transitions'', hep-th/9801205 ;
        O. DeWolfe and B. Zwiebach, ``String junctions for arbitrary
        Lie algebra representations'', hep-th/9804210;
        T. Hauer, ``Equivalent string networks and uniqueness of
        BPS states'', hep-th/9805076  ;
        O. DeWolfe, T. Hauer, A. Iqbal and B. Zwiebach, ``Constraints
        on the BPS spectrum of N=2, D=4 theories with A-D-E
        flavor symmetry'', hep-th/9805220 .

\bibitem{rey} S-J. Rey and J-T. Yee, ``BPS dynamics of triple 
        (p,q) string junction'', hep-th/9711202.

\bibitem{lee} M. Krogh and S. Lee, ``String network from M
        theory'', hep-th/9712050.

\bibitem{mats} Y. Matsuo and K. Okuyama, ``BPS condition on
        string junction from M theory'', hep-th/9712070.

\bibitem {berg} O. Bergman, ``Three pronged strings and 
        1/4 BPS states in N=4 super-Yang-Mills theories''
        hep-th/9712211;
        O. Bergman and A. Fayyazuddin, ``String junctions
        and BPS states in Seiberg-Witten theory'', hep-th/9802033.

\bibitem{bhatt} S. Bhattacharyya, A. Kumar and S. Mukhopadhyay,
        ``String network and U-duality'', 
	{\prl} {\bf 82} (1998) 754, [hep-th/9801141]. 
     
 
\bibitem{cal} C. G. Callan and L. Thorlacious, ``World sheet 
        dynamics of string junctions'', hep-th/9803097.


\bibitem{rest} K. Hasimoto, H, Hata and N. Sasakura,``Three-string
        junction and BPS saturated solutions in SU(3) supersymmetric
        Yang-Mills theory'', hep-th/9803127;
        ``Multipronged strings and BPS saturated solutions in SU(N)
        supersymmetric Yang-Mill's theory'', hep-th/9804164;
        T. Kawano and K. Okuyama, ``String network and BPS 
        saturated solutions in SU(3) supersymmetric Yang-Mills 
        theory'', hep-th/9803127; 
        K. Li and S. Yi, ``Dyons in N=4 supersymmetric theories
        and 3-pronged string'' , hep-th/9804174;

\bibitem{bk} O. Bergman and B. Kol, ``String webs and 
        1/4 BPS monopoles'', hep-th/9804160.

\bibitem{rovel} C. Rovelli, ``Loop Quantum Gravity'' 
        gr-qc/9710008.

\bibitem{smo2} L. Smolin, ``Strings as perturbations of evolving spin-networks
 hep-th/9801022. 

\bibitem{smolin} F. Markopoulou and L. Smolin, ``Non-perturbative
        dynamics of abstract (p,q) string networks'', 
        hep-th/9712148.

\bibitem{wit} A. Hannany and E. Witten, ``Type IIB superstrings,
        BPS monopoles, and Three-dimensional gauge dynamics'', 
        {\np}{\bf 492} (1997) 152, hep-th/9611230   ;
        E. Witten, ``Solutions of four dimensional fields theories
        via M theory'', {\np}{\bf 500} (1997), 3, hep-th/9703166.
\bibitem{aha} O. Aharony and A. Hanany, ``Branes, 
        superpotentials and superconformal fixed points'', 
        {\np}{\bf 504} (1997) 239 hep-th/9704170;
        O. Aharony, A. Hanany and B.Kol, ``Webs 
        of (p, q) five-branes, five-dimensional theories and
        grid diagrams'', {\it JHEP} {\bf 01} (1998), hep-th/9710116.

\bibitem{leung} N. Leung and C. Vafa, ``Branes and toric
        geometry'', hep-th/9711013.

\bibitem{cvet} M. Cvetic and C. Hull, {\np}{\bf 480} (1996) 296, 
	[hep-th/9606193].

\bibitem{pope} E. Cremmer, H. Lu, C. Pope and K. Stelle, 
	{\np} {\bf 520} (1998) 132, [hep-th/9707207];
	H. Lu, C. Pope, T. Tran and K. Xu, {\np}
	{\bf 511} (1998) 98, [hep-th/9708055]. 

\bibitem{malda} S. Ferrara and J. Maldacena, hep-th/9706097;
        L. Andrianopoli, R.D'Auria and S. Ferrara, ``U-Invariants,
        Black Hole Entropy and Fixed Scalars'', hep-th/9703156. 

\bibitem{vafa} C. Vafa, {\np}{\bf 469} (1996) 403 [hep-th/9602022];
        A. Kumar and C. Vafa, ``U-manifolds'', {\pl}{396} 
        (1995) 85, [hep-th/9611007].

\bibitem{ubrane} J. Liu and R. Minasian, ``U-branes and $T^3$
        fibrations'', hep-th/9707125

\bibitem{sl2} J. Schwarz, ``An SL(2,Z) multiplet of type 
        IIB superstring'', {\pl}{\bf 360} (1995) 13, 
        [hep-th/9508143].

\bibitem{bbg} G. W. Gibbons, M. B. Green and M. J. Perry,
        ``Instantons and seven branes in type IIB superstring 
        theory'', {\pl}{\bf 370} (1996) 37, hep-th/9511080 ;
        B. R. Greene, A. Shapere, C. Vafa and S-T Yau,
        ``Stringy cosmic strings and non-compact Calabi-Yau
        manifolds'', {\np}{\bf 337} (1990) 1.
\bibitem{tani} Y. Tanii, ``N=8 supergravity in six dimensions'',
        {\pl}{\bf 145} (1984) 197.

\bibitem{ahaha} O. Aharony, J. Sonnenschein and 
        S. Yankielowicz, ``Interactions of strings 
        and D-branes from M theory'' 
        {\np}{\bf 474} (1996) 309, [hep-th/9603009]. 

\bibitem{ss} E. Cremmer, ``Supergravities in 5 dimensions''
        in {\it Superspace and Supergravity}, Eds. S. W. Hawking and
        M. Rocek, Cambridge University Press, 1981.

\bibitem{nonbps} A. Sen, ``Stable non-BPS states in string theory'', 
        hep-th/9803194; ``Stable Non-BPS Bound States of BPS
        D-branes'', hep-th/9805019. 

\bibitem{lu} J. X. Lu and S. Roy, ``U Duality P-branes in diverse
        dimensions'', hep-th/9805180.



\end{thebibliography}
\end{document}